**The Journal Impact Factor Should Not Be Discarded**

Running title: JIF Should Not Be Discarded

Lutz Bornmann,[1] Alexander I. Pudovkin[2]

[1]Division for Science and Innovation Studies, Administrative Headquarters of the Max Planck Society, Munich, Germany

[2]A. V. Zhirmunsky Institute of Marine Biology, Far Eastern Branch of Russian Academy of Sciences, Vladivostok, Russian Federation

Address for Correspondence: Lutz Bornmann, PhD

Division for Science and Innovation Studies, Administrative Headquarters of the Max Planck Society,

Hofgartenstraße 8, 80539 München, Germany

E-mail: bornmann@gv.mpg.de

Journal editors and experts in scientometrics are increasingly concerned with the reliability of the Journal Impact Factor (JIF, Clarivate Analytics, formerly the IP & Science business of Thomson Reuters) as a tool for assessing the influence of scholarly journals. A paper byLarivière et al. (1), which was reposited on *bioarXiv* portal and commented on in *Nature* (2), reminded all stakeholders of science communication that the citability of most papers in an indexed journal deviates significantly from its JIF. These authors recommend to display journal citation distribution instead of the JIF, and the proposal is widely discussed on social networking platforms (3,4).

The overall impression is that the discussion over the JIF is endless. The JIF along with the h-index is the simplest and most studied indicator in scientometrics (5,6). However, the commentary in *Nature* (2) and subsequent debates over the citation distribution revived interest of the scientific community toward empirical analyses of the JIF and its uses and misuses in research evaluation.

After all the endless discussions, research evaluators should have realized that the JIF should not be used to measure the impact of single papers. But there are still some experts, who argue that the use of the JIFs at the level of single papers cannot be simply distinguished from its use at the journal level (4). In some circumstances, the JIFs may help authors and readers to pick, read, and cite certain papers. Papers from high-impact journals are more likely to be picked and cited than similar ones from low-impact periodicals.

The JIF should not be demonized. It still can be employed for research evaluation purposes by carefully considering the context and academic environment. Elsevier – provider of the Scopus database – rates the JIF as so important that the company introduced the near-doppelgänger CiteScore, recently (see https://journalmetrics.scopus.com/). The JIF measures the average impact of papers, which are published in a journal, with a citation window of only one year. The JIFs are calculated and published annually in the Journal Citation Reports (JCR, Clarivate Analytics). Papers counted in the denominator of the JIF formula are published within 2 years prior to this citation metric calculation. In contrast to the JIF, the new CiteScore metric considers the papers from 3 years (instead of 2 years). As such, the JIF (and also the CiteScore) covers rather short term of interest toward papers (i.e., interest at the research front) and overlooks long-term implications of publication activity (the so-called sticky knowledge) (7). Focus on the short-term attention of the field-specific community makes sense since the JIF was initially designed to guide librarians purchase the most used modern periodicals for their libraries. Accordingly, the JIF cannot and should not be employed for evaluating the average impact of papers in a journal in the long and distant run.

The JIF formula aims at calculating average numbers that reveal the central tendency of a journal's impact. As such, one or few highly-cited papers, which are published within the 2 years, may boost the JIF. That is particularly the case with *Nature*, *Science*, and other influential journals (1). The skewed citation distribution implies that the JIF values do not reflect the real impact of most papers published in the index journal. The absolute number of citations received by a single paper is the correct measure of its impact. Currently, the Web of Science and Scopus databases can provide an outlook at citations for evaluating the impact of single papers.

Importantly, the JIF is the best predictor of single papers' citability (8). Studies examining the predictive value of the JIF along with number of authors and pages proved that notion (9). One can expect more citations to single papers, which are published in higher-impact journals, compared to those in lower-impact ones.

Another important point is the field-dependency of citations contributing to the JIFs. There are differing citation rates across different disciplines and subject categories, regardless of the scientific quality of the papers, and confounded by field-specific authorship rules, publication activity, and referencing patterns (10). Such differences justified the development of field-normalized indicators, which are employed for evaluating individual researchers, research groups, and institutions (11,12). Since the JIF is not a field-normalized indicator, it can only be used for evaluations within a single subject category.

The SCImago Journal Rank SJR indicator – a variant of the JIF – were employed for institutional excellence mapping at www.excellencemapping.net (13,14). For institutions worldwide, this site maps the results of 2 indicators. First, the 'best paper rate' measures the long-term impact of papers in a size-independent way, using percentiles as a field-normalized indicator. Second, the 'best journal rate,' which is based on the citation impact of the journals publishing the institutions' papers. That indicator is the proportion of papers published in journals belonging to the 25% of the best journals in their subject categories in terms of citation impact. Through the consideration of journal sets, the indicator is a field-normalized metric at the journal level. The indicator demonstrates how successful are academic institutions in terms of publishing their papers in high-impact journals (13,14). Thus, the so-called success at www.excellencemapping.net is measured by the ability of publishing in high-impact target journals and by receiving the long-term attention of the scientific community.

The JIF can be used to measure the ability of individual researchers and institutions to successfully publish their research. However, the JIF should not be used as a proxy for measuring the impact of single papers. In this regard, more appropriate indicators should be considered (e.g., data from the "Field Baselines" tables in the Essential Science Indicators [ESI] by Clarivate Analytics). The baselines can be used to assess whether a specific paper received an impact which is far above or below the worldwide average performance in a field. For example, the 2006 baseline for chemistry is approximately 23 (November 14, 2016). If a chemistry paper from 2006, which is published by an evaluated entity, attracts 50 citations, the impact of that paper is far above the baseline, whereas with 10 citations the impact would be far below the baseline.

There is only one scenario when the use of the JIFs is justifiable for the assessment of individual scientists (15). It is when recently published papers are considered for research evaluation, which is routinely practised for intramural monitoring of staff productivity, academic promotion, or recruitment. The evaluators pay particular attention to the most recent publications. But for these items, the citation window is too short for quantifying their citation impact reliably (16), and in that case reputation of the publishing journals along with their JIFs can be conditionally employed as the proxies of single papers' impact (9). InCites (Clarivate Analytics) has already implemented the calculation of specialty-specific percentile-transformed JIFs (17), which reflect field-normalized journal impact values and can be used for assessing recently published papers.

**DISCLOSURE**

The authors have no potential conflicts of interest to disclose.

**AUTHOR CONTRIBUTION**

Study design: Bornmann L, Pudovkin AI. Writing: Bornmann L, Pudovkin AI. Revision and final approval of manuscript: all authors.

**ORCID**

Lutz Bornmann http://orcid.org/0000-0003-0810-7091